\documentclass[aps,prc,twocolumn,groupaddress,showkeys]{revtex4-1}
\pdfoutput=1
\usepackage{graphicx} 
\usepackage{dcolumn} 
\usepackage{bm}
\usepackage{color}
\usepackage{hyperref}
\usepackage{graphicx}
\usepackage{amsmath}
\usepackage{setspace}

\begin{document}

\title{Initialization effects of nucleon profile on the $\pi$ yields in heavy-ion collisions at medium energies}
\author{Zu-Xing Yang$^{1,2}$}
\email[]{yangzuxing16@impcas.ac.cn}
\author{Nicolas Michel$^{1}$}
\email[]{nicolas.michel@impcas.ac.cn}
\author{Xiao-Hua Fan$^{3}$}
\author{Wei Zuo$^{1,2}$}
\affiliation{$^1$Institute of Modern Physics, Chinese Academy of Sciences, Lanzhou 730000, China\\
$^2$School of Nuclear Science and Technology, University of Chinese Academy of Sciences, Beijing 100049, China\\
$^3$School of Physical Science and Technology, Southwest University, Chongqing 400715, China
}

\begin{abstract}

We study a problem of $\pi$ production in heavy ion collisions in the context of the Isospin-dependent Boltzmann-Uehling-Uhlenbeck (IBUU) transport model.
We generated nucleon densities using two different models, the Skyrme-Hartree-Fock (SHF) model and configuration interaction shell model (SM). 
Indeed, inter-nucleon correlations are explicitly taken into account in SM, while they are averaged in the SHF model. 
As an application of our theoretical frameworks, we calculated the $\pi^{-}$ and $\pi^{+}$ yields in collisions of nuclei with $A = 30-40$ nucleons.
We used different harmonic oscillator lengths $b_{HO}$ to generate the harmonic oscillator basis for SM in order to study both theoretical and experimental cases.
It is found that SM framework with $b_{HO}$ = 2.5 fm and SHF can be distinguished by the yield of $\pi$ mesons, in this case the density distribution calculated by the shell model produces more $\pi$ in the collision. 
In comparison, SM with $b_{HO}$ = 2.0 fm is characterized from SHF by the double $\pi^{-}/\pi^{+}$ ratios with different large impact parameters, from which one can find the double $\pi^{-}/\pi^{+}$ ratios of SM change smoother and are less than those of SHF.  

\end{abstract}

\maketitle

\section{Introduction}

Research on heavy-ion collisions (HIC) has been extensively performed experimentally and theoretically over the past few decades \cite{BUUguide}. 
While their theoretical description has relied on phenomenology for a long time \cite{ BUUguide2}, microscopic models have been developed in order to have a more realistic approach of HIC \cite{BUUguide,BUUguide2,BUUguide3}.
It has been shown recently that inclusion of pairing correlations can significantly modify reaction observables such as counts of emitted protons in proton-target reactions at intermediate energies by exploring the density profile of target nuclei\cite{SHF20194}. 
Deformation and orientation effects have also been noticed to be important for particle production in uranium-uranium collisions at relativistic energies \cite{BUUden}.
Consequently, both structure and reaction dynamics are important for the description of nucleus-nucleus collisions.

Recent experimental data consist of the $^{197}$Au +$^{197}$Au reaction at 400 MeV/nucleon for incident energy within the ASY-EOS experiment at the GSI laboratory \cite{EXP1},
and of the $^{132}$Sn + $^{132}$Sn reaction at a beam energy of 0.3 GeV/nucleon, carried out at RIKEN in Japan \cite{EXP2}.
They could provide information about the equation of state (EOS) \cite{BUUeos,BUUpieos,BUUeos2,BUUeos3}, short-range correlations \cite{BUUsrc} and medium effects in scattering cross sections \cite{BUUcross,BUUpi}.
Both theoretical and experimental studies have pointed out that the $\pi$ meson is one of the most promising probes to study the nuclear structure and reaction aspects of heavy ion collisions \cite{BUUpieos,BUUeos2,BUUpi,BUUpi2}.
In order to calculate the $\pi$ yields produced during the collisions of medium and heavy nuclei, the Isospin-dependent Boltzmann-Uehling-Uhlenbeck (IBUU) transport model is often used.
The importance of $\pi$ mesons for constraining the EOS of asymmetric nuclear matter was firstly pointed out by Bao-An Li et al \cite{BUUpi2}.
In the case of incident energy near the $\pi$ meson threshold, it happens that the $\pi$ meson is essentially generated by the decay of the $\Delta_{1232}$ resonant state, whereas the $N^{*}$ resonant state can be ignored \cite{BUUguide,BUUpi2}. 
In recent years, Stone et al have studied systematically the effects of the proton and neutron density distributions in central heavy-ion collisions, and found that the maximal neutron-to-proton ratio during the collision process is quite sensitive to the initial-state density distributions\cite{BUUpi3}, where the different maximal neutron-to-proton ratios would result in difference of $\pi$ productions.
Therefore, it was also shown recently that $\pi$ mesons provide a remarkably good probe to study bubble structure inside nuclei \cite{buuepja}. 

In fact, the Skyrme-Hartree-Fock (SHF) model\cite{shfori, shfori2}, used along with density functional theory, is one of the most efficient and widely used models to study the bulk properties of nuclei all over the nuclear chart \cite{shfori,SHF20191,SHF20193}.
As the contribution of the wave function provided by the mean-field Slater determinant describing nuclear ground states does not exceed 70\%, typically, it is necessary to use models beyond mean-field to account more realistically for the effect of many-body interactions in nuclear wave functions \cite{SM2005}.
The aim of this article is then to accurately simulate experimental situations with the SHF and SM approaches, in order to identify the main effects of inter-nucleon correlations on $\pi$ yields.
In order to include the main effects of inter-nucleon correlations, while dealing with a numerically tractable model, we chose to use the core + valence nucleons shell model (SM) approach \cite{SM2005,SM20052,SM20053,SM20054}.
Both SHF and SM approaches will be used to evaluate nucleon densities, which are then adopted in the IBUU model to calculate $\pi$ yields in HIC.
We will consider nuclei of $sd$-shell, bearing $A=30-40$ nucleons.
Indeed, SHF can be reliably used in these nuclei due to their sizable number of nucleons, while SM leads to tractable shell model spaces in the considered region of nuclear chart.
The Hamiltonians used consist of the SkM$^*$ interaction in SHF\cite{SHF20194,skm*},
while the USDB interaction will be considered as SM interaction (which one will denote as USDB-SM), as it has proved to properly describe the properties of nuclei of the $sd$-shell\cite{USDB}.



\section{The IBUU transport model}

The phase-space distribution function $f(\vec{r}, \vec{p}, t)$ of nucleons is needed to calculate the $\pi$ yields in HIC. 
In the IBUU model\cite{BUUpi2} the time-evolution of $f(\vec{r}, \vec{p}, t)$ is described by the following transport equations:
\begin{equation}
\frac{\partial f}{\partial t}+\nabla_{\vec{p}} E \cdot \nabla_{\vec{r}} f-\nabla_{\vec{r}} E \cdot \nabla_{\vec{p}} f=I_{c},
\label{buu001}
\end{equation}
where $E$ is the single particle energy which contains the kinetic energy $E_{kin}$ and potential energy $U$, and $I_c$ is the so-called collision item, embedding the modification of phase-space distribution function
induced by elastic and inelastic two-body collisions \cite{BUUpi}.
Thus, the left and right sides of Eq.(\ref{buu001}) represent the time evolution and collision effects involving a single particle in the nuclear mean-field, respectively.

In this model, the momentum-dependent single nucleon potential $U$ is a function of nucleon densities and reads \cite{BUUpi,yongprc2016}:
\begin{eqnarray}
U(\rho,\delta,\vec{p},\tau)&=&A_u(x)\frac{\rho_{\tau'}}{\rho_0}+A_l(x)\frac{\rho_{\tau}}{\rho_0}\nonumber\\
& &+B(\frac{\rho}{\rho_0})^{\sigma}(1-x\delta^2)-8x\tau\frac{B}{\sigma+1}\frac{\rho^{\sigma-1}}{\rho_0^\sigma}\delta\rho_{\tau'}\nonumber\\
& &+\frac{2C_{\tau,\tau}}{\rho_0}\int
d^3\,{p^{'}}\frac{f_\tau(\vec{r},\vec{p^{'}})}{1+(\vec{p}-\vec{p^{'}})^2/\Lambda^2}\nonumber\\
& &+\frac{2C_{\tau,\tau'}}{\rho_0}\int
d^3\,{p^{'}}\frac{f_{\tau'}(\vec{r},\vec{p^{'}})}{1+(\vec{p}-\vec{p^{'}})^2/\Lambda^2},
\label{buupotential}
\end{eqnarray}
where $\tau, \tau'=1/2(-1/2)$ is the isospin projection of neutrons (protons), $\delta=(\rho_n-\rho_p)/(\rho_n+\rho_p)$ denotes the isospin asymmetry, and $\rho_n$, $\rho_p$ denote neutron and proton number densities, respectively. 
The parameters $x , A_u(x), A_l(x), B, C_{\tau,\tau}, C_{\tau,\tau'}, \sigma, \Lambda$ are standard in the IBUU model and their definition can be found in Ref.\cite{yongprc2016}. 
The value $x=1$ is used for the soft symmetry energy parameter $x$ in this work.
 
The following formula is used for the $\Delta$ resonance potential:
\begin{equation}
\begin{array}{l}{U^{\Delta^{-}}=U_{n}, \quad U^{\Delta^{0}}=\frac{2}{3} U_{n}+\frac{1}{3} U_{p}}, \\ {U^{\Delta^{+}}=\frac{1}{3} U_{n}+\frac{2}{3} U_{p}, \quad U^{\Delta^{++}}=U_{p}}\end{array}.
\end{equation}
The effective masses of neutron, proton and  $\Delta$ resonance are directly deduced from their corresponding potential, i.e.,
\begin{equation}
\frac{m_{B}^{*}}{m_{B}}=1 /\left(1+\frac{m_{B}}{p} \frac{d U}{d p}\right).
\end{equation}

One obtains the isospin-dependent baryon-baryon (BB) scattering cross section in medium, $\sigma_{B B}^{\text {medium }}$:
\begin{equation}
\begin{aligned} R_{\text {medium }}^{B B}(\rho, \delta, \vec{p}) & \equiv \sigma^{\text{medium}}_{B B_{\text {elastic, inelastic }}} / \sigma_{B B_{\text {elastic, inelastic }}}^{\text{free}} \\ &=\left(\mu_{B B}^{*} / \mu_{B B}\right)^{2} \end{aligned},
\label{eq_BB}
\end{equation}
where $\mu_{BB}$ and $\mu_{BB}^*$ are the reduced masses of the colliding baryon pairs in free space and medium, respectively, and $\sigma_{B B_{\text {elastic, inelastic }}}^{\text{free}}$ is the free BB scattering cross section.
The elastic proton-proton cross section $\sigma_{pp}$ and neutron-proton cross section $\sigma_{np}$ are taken from experimental data, and $\sigma_{nn}$ is assumed to be equal to $\sigma_{pp}$. 
The N$\Delta$ free elastic cross sections are assumed to be equal to nucleon-nucleon (NN) elastic cross sections at the same center of mass energy.
The inelastic NN cross sections read:
\begin{equation}
  \begin{aligned} \sigma^{pp \rightarrow n \Delta^{++}} &=\sigma^{n n \rightarrow p \Delta^{-}}=\sigma_{10}+\frac{1}{2} \sigma_{11} \\ \sigma^{p p \rightarrow p \Delta^{+}} &=\sigma^{n n \rightarrow n \Delta^{0}}=\frac{3}{2} \sigma_{11} \\ \sigma^{n p \rightarrow p \Delta^{0}} &=\sigma^{n p \rightarrow n \Delta^{+}}=\frac{1}{2} \sigma_{11}+\frac{1}{4} \sigma_{10} \end{aligned},
  \label{eq_NN}
\end{equation}
and are parametrized via (see Ref.\cite{BUUguide}):
\begin{equation}
\sigma_{I I^{\prime}}(s)= \frac{\pi(\hbar c)^{2}}{2 p^{2}} \alpha\left(\frac{p_{r}}{p_{0}}\right)^{\beta} \frac{m_{0}^{2} \Gamma^{2}\left(q / q_{0}\right)^{3}}{\left(s^{*}-m_{0}^{2}\right)^{2}+m_{0}^{2} \Gamma^{2}},
  \label{eq_NN_parameters}
\end{equation}
where: 
\begin{eqnarray*}
  s^{*}&=&\langle M\rangle^{2}, \\
 p_{r}^{2}(s)&=&\frac{\left[s-\left(m_{N}-\langle M\rangle\right)^{2}\right]\left[s-\left(m_{N}+\langle M\rangle\right)^{2}\right]}{4 s}, \\
q^{2}\left(s^{*}\right)&=&\frac{\left[s^{*}-\left(m_{N}-m_{\pi}\right)^{2}\right]\left[s^{*}-\left(m_{N}+m_{\pi}\right)^{2}\right]}{4 s^{*}},  \\
  q_{0}&=&q\left(m_{0}^{2}\right),
  \end{eqnarray*}
and $I$,$I^{\prime}$ represent the isospins of the initial and final states of the two nucleons, respectively.
As for $\alpha, \beta, m_{0},$ and $\Gamma$, there are four parameter sets for $\sigma_{10}^{d},  \sigma_{11},  \sigma_{10},  \sigma_{01}$.
Details concerning the value $\langle M\rangle$ and the four $\sigma$ parameter sets can be found in Ref.~\cite{VerWest}. 
  
The mass of the produced $\Delta$ baryons in Eqs.(\ref{eq_BB},\ref{eq_NN},\ref{eq_NN_parameters}) is described by a modified Breit-Wigner function \cite{lba2001}.	
\begin{equation}
P\left(m_{\Delta}\right)=\frac{p_{f} m_{\Delta} \times 4 m_{\Delta_0}^{2} \Gamma_{\Delta}}{\left(m_{\Delta}^{2}-m_{\Delta_0}^{2}\right)^{2}+m_{\Delta_0}^{2} \Gamma_{\Delta}^{2}},
\label{buu002}
\end{equation} 
where $m_{\Delta_0}$ is the centroid of the resonance, $\Gamma_{\Delta}$ is the width of the resonance and $p_{f}$ is the center of mass momentum in the N$\Delta$ channel.

As for the reaction cross section of the reverse reaction, Danielewicz and Bertsch first deduced and obtained the isospin-averaged cross section in Ref.\cite{eq9}:
\begin{eqnarray}
\sigma_{N \Delta \rightarrow N N} &=& \left( \frac{m_{\Delta} p_{f}^{2} \sigma_{N N \rightarrow N \Delta}}{2(1+\delta) p_{i}} \right) \nonumber \\
&\times& \left( \int_{m_{\pi}+m_{N}}^{{\sqrt{s}-m_{N}}} \frac{d m_{\Delta}}{2 \pi} P \left( m_{\Delta} \right) \right)^{-1},
\end{eqnarray} 
where $p_f$ and $p_i$ are the nucleon center of mass momenta in the NN and N$\Delta$ channels, respectively. The width of $\Delta$ resonance is given by:
\begin{equation}
\Gamma_{\Delta}=\frac{0.47 q^{3}}{m_{\pi}^{2}\left[1+0.6\left(q / m_{\pi}\right)^{2}\right]},
\end{equation}
where $q=\sqrt{\left(\frac{m_{\Delta}^{2}-m_{n}^{2}+m_{\pi}^{2}}{2 m_{\Delta}}\right)^{2}-m_{\pi}^{2}}$, and $q$ is the $\pi$ momentum in the $\Delta$ rest frame. 
For the reaction $\Delta\rightarrow\pi+N$, the classical formula $P_{\text {decay }}=1-\exp \left(-d ~ t ~ \Gamma_{\Delta} / \hbar\right)$ is used for the decay probability of the $\Delta$ particle.
As before, a Breit-Wigner function is assumed for the $\pi + N$ cross section in the reverse reaction  \cite{lba2001,eq9}:
\begin{equation}
\sigma_{\pi+N}=\sigma_{\max }\left(\frac{q_{0}}{q}\right)^{2} \frac{\frac{1}{4} \Gamma_{\Delta}^{2}}{\left(m_{\Delta}-m_{\Delta 0}\right)^{2}+\frac{1}{4} \Gamma_{\Delta}^{2}},
\end{equation}
where $q_{0}$ is the $\pi$ momentum  at the centroid of the resonance  $m_{\Delta_0}= 1.232 ~\text{GeV}$.

The maximal cross section $\sigma_{\max }$ reads as follows:
\begin{equation}
\begin{aligned} \sigma_{\max }^{\pi^{+} p \rightarrow \Delta^{++}} &=\sigma_{\max }^{\pi^{-} n \rightarrow \Delta^{-}}=200 \mathrm{mb} \\ \sigma_{\max }^{\pi^{-} p \rightarrow \Delta^{0}} &=\sigma_{\max }^{\pi^{+} n \rightarrow \Delta^{+}}=66.67 \mathrm{mb} \\ \sigma_{\max }^{\pi^{0} p \rightarrow \Delta^{+}} &=\sigma_{\max }^{\pi^{0} n \rightarrow \Delta^{0}}=133.33 \mathrm{mb} \end{aligned}.
\end{equation}
All elastic scattering and decay reactions are assumed to be isotropic in this model. Elastic scattering cross sections bear an angular distribution \cite{Cugnon1981}: 
\begin{equation}
\frac{d \sigma_{\mathrm{el}}}{d \Omega} \propto e^{b t},
\end{equation}
where: $$b=\frac{6[3.65(\sqrt{s}-1.8766)]^{6}}{1+[3.65(\sqrt{s}-1.8766)]^{6}}$$ and $t=-2 p^{2}(1-\cos \theta)$, with $p$ is the momentum of one particle in the center of mass frame.

Baryon-baryon collisions in the IBUU model are simulated by a Monte Carlo-based modelling process \cite{BUUguide}.
During a small time interval ($\delta t = 0.5$ fm/c), one firstly checks if two baryons are sufficiently close to interact with each other ($b_{\max }=\sqrt{\sigma_{\mathrm{nn}}^{t}(\sqrt{s}) / \pi}$).
If it is the case, one performs simulations to determine which kind of reaction should be present \cite{BUUguide}.
When a scattering reaction takes place between two nucleons, both elastic and inelastic scattering processes can arise. NN elastic reactions occur with the probability of collision $P_{elastic}(NN) = \sigma_{elastic} /\sigma_{\mathrm{nn}}^{t}(\sqrt{s}) $.
$P_{inelastic}(NN \rightarrow N\Delta) = \sigma_{inelastic} /\sigma_{\mathrm{nn}}^{t}(\sqrt{s})$ determines the probability of baryon pair formation in the $NN \rightarrow N\Delta$ reaction, with $\sigma_{\mathrm{nn}}^{t}(\sqrt{s})$ the total cross section in the center-of-mass energy $\sqrt{s}$, and $P_{elastic}+P_{inelastic}=1$.
For the baryon $\Delta$, the same procedure has been performed in our simulations, which has three branches: 

i. elastic scattering with the nucleon ($N\Delta$); 

ii. inelastic scattering with the nucleon to produce two nucleons ($N\Delta\rightarrow NN$); 

iii. decay to a $\pi$ meson and a nucleon ($\Delta\rightarrow N\pi$). 

The case of other baryons and mesons are  similar and we won't go into further detail.
The present results have been obtained using the IBUU04 program\cite{BUUguide,SHF20194,BUUeos,BUUpieos,BUUeos2,BUUsrc,BUUpi}, using as inputs the nucleon densities calculated with SHF and SM, respectively.
$\pi$ yields are then determined from the Monte Carlo process described above and coded in the IBUU04 program.

\section{Nucleon densities in the SHF and USDB-SM approaches}

The density distributions of the two colliding nuclei used in the IBUU model (see Eq.(\ref{buupotential})) are isotropic.
Consequently, only radial densities are needed to be calculated from the SHF and USDB-SM approaches.
Only valence nucleons are active in the USDB-SM approach. 
Therefore, approximations are needed to evaluate a function equivalent to the SHF nucleon density distribution.
For this, one firstly calculates the total density of $^{\rm A}\rm S (A=32, 34, 36)$, $^{\rm A}\rm Cl (A=35, 37)$ and $ ^{16}\rm O$ with SHF.
Nucleon distributions in a SM approach are then defined by adding on the density distribution of $ ^{16}\rm O$ from the valence nucleon density distribution of the considered nucleus:
\begin{equation} \label{SHF1}
\rm \rho_r^{(n,p)}( ^A X)=\rho_r^{(n,p)}({^AX(v.n.)})+ \rho_r^{(n,p)}(^{16}O)
\end{equation}
\noindent where X=(S,Cl) and (v.n.) signifies valence nucleons.


In our SM calculations, different harmonic oscillator length $b_{HO} =$ 1/$\alpha$ = $ \sqrt{\hbar/\mu\omega_0}$ are used.
On the one hand, in the SHF model, the length $b_{HO}$ = $1.42 \times A^{1/6} \approx $ 2.5 fm is considered.
Hence, it is for this value that the distributions issued from USDB-SM and SHF have closest maxima for valence nucleons, so that it is physically sound to use $b_{HO}$ = 2.5 fm in SM for our purpose.
On the other hand, RMS nuclear charge radii play a crucial role in experimental observations.
Thus, to fit the RMS of experimental nuclear charge radii in USDB-SM, we used another $b_{HO}$ value, which is $b_{HO}$ = 2 fm. 
For brevity, we will designate the case USDB-SM ($b_{HO}$ = 2 fm) by SM1 and that of USDB-SM ($b_{HO}$ = 2.5 fm) by SM2 in the following.
The experimental (EXP) and calculated RMS nuclear charge radii in the different used models (SHF, SM1 and SM2) are shown in Table.\ref{tabR}.
One can see in Table.\ref{tabR} that SM1 reproduces experimental RMS nuclear charge radii along with SHF, whereas those provided by SM2 are too large by about 0.5 fm.

\begin{table}[h]
\renewcommand{\arraystretch}{1.2}
        \caption{RMS nuclear charge radii (in fm). See text for the definition of acronyms.}
\begin{tabular}{@{}lllll@{}}
\toprule
     & SHF   & SM1   & SM2   & EXP   \\
\toprule
~~~$^{32}$S~~~~~~~  & 3.256~~~~~~~ & 3.282~~~~~~~ & 3.804~~~~~~~ & 3.261~~~ \\
\hline
~~~$^{34}$S  & 3.276 & 3.282 & 3.803 & 3.285 \\
\hline
~~~$^{36}$S  & 3.300 & 3.282 & 3.803 & 3.298 \\
\hline
~~~$^{35}$Cl & 3.328 & 3.310 & 3.857 & 3.365 \\
\hline
~~~$^{37}$Cl & 3.348 & 3.310 & 3.856 & 3.384 \\
\toprule

\end{tabular}
\label{tabR}
\end{table}

\begin{figure}
\includegraphics[width=9 cm]{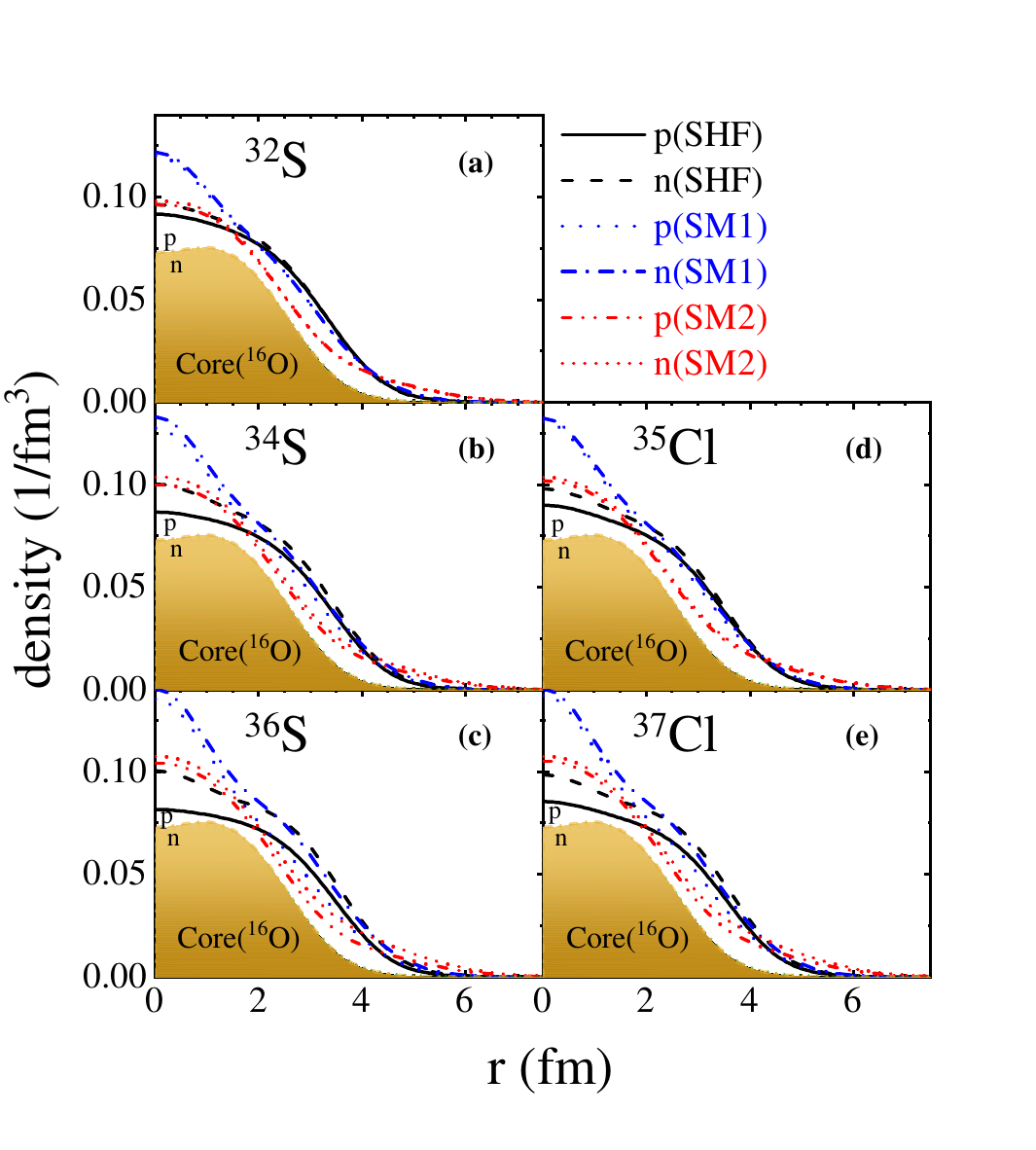}
\caption{\label{fig:init_S} (Color online)
  Density distributions of sulfur isotopes (a)-(c) and chlorine isotopes (d)-(e) obtained with SHF, SM1 and SM2.
  The normalization condition $4\pi\int_{0}^{r_{\max }} \rho(r) r^{2} d r=A$ is used, with $A$ the number of nucleons.
  The density distributions of protons and neutrons (a)-(e) of the $^{16}\rm O$ core are also shown in the context of SHF,
  using the normalization condition $4\pi\int_{0}^{r_{\max }} \rho(r) r^{2} d r=A_{core}$.}
\end{figure}

\begin{figure}
\includegraphics[width=9 cm]{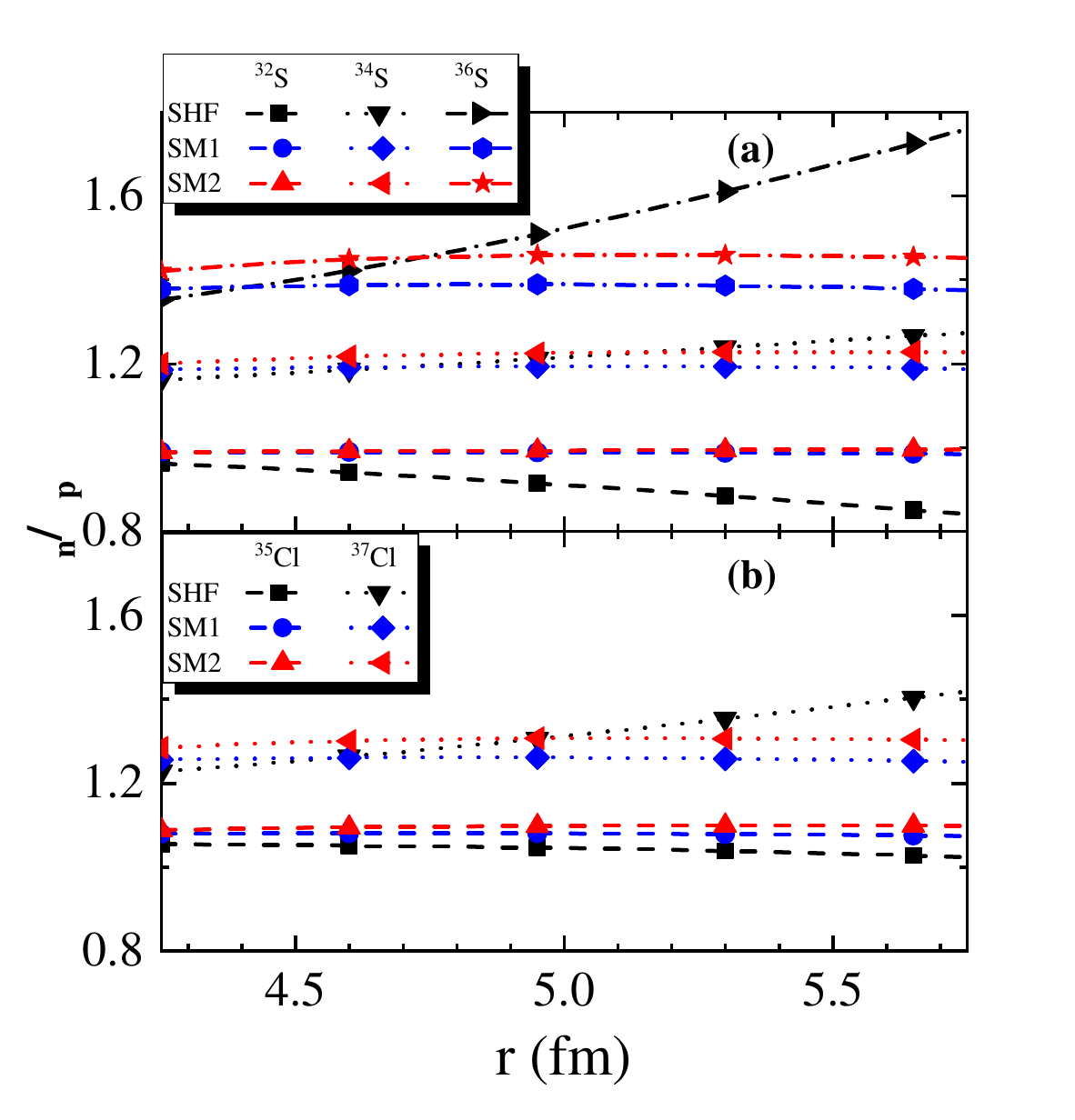}
\caption{\label{fig:npratio_r} (Color online) Ratio of neutron density to proton density as a function of radius in the sulfur (a) and chlorine isotopic chain (b), using the USDB-SM and SHF approaches.}
\end{figure}

Fig.\ref{fig:init_S} shows the density distributions of different nuclides obtained by SHF, SM1 and SM2 with the core $\rm ^{16}O$ obtained by SHF.
The calculated density distributions are generated almost entirely by valence nucleons for radii larger than 4.5 fm.
One can see that the extension of USDB-SM density distributions is larger than those arising from SHF, especially for SM2.
Because the configuration mixing gives nucleons a higher chance of occupying higher orbits than meaning filed method.
This is exactly how the inter-nucleon correlations work.
Fig.\ref{fig:npratio_r} shows the $\rm \rho_n/\rho_p$ ratio as a function of radius  for different nuclides, using the USDB-SM and SHF approaches.
the ratios of neutron density to proton density in SHF and USDB-SM are close for radii around 4.5 fm. 
As the radius increases to about 5.5 fm, SM1 and SM2 results still remain close, while the difference between SHF and USDB-SM results increases significantly.
This result is also due to the inter-nucleon correlations.
Take $\rm ^{36}S$ as an example:
In SHF model, When the radius is greater than 5 fm, the density distribution of protons is sharply reduced, which leads to an increase in $\rm \rho_n/\rho_p$.
For USDB-SM, both proton and neutron distributions are extended, which makes the variation of $\rm \rho_n/\rho_p$ smoother.
One can see, at larger radii, the $\rm \rho_n/\rho_p$ ratio in USDB-SM closer to the $n/p$ ratio of the system than that in SHF model.

In the present paper, we shall consider the following nuclear ground states: $\rm ^{32}S(0^+)$,  $\rm^{34}S(0^+)$,  $\rm^{36}S(0^+)$, $\rm^{35}Cl(3/2^+)$ and $\rm^{37}Cl(3/2^+)$. 
The $^{16}$O core is then not expected to play a significant role in $\pi$ production.
We will consider that the reactions of interest have a large impact parameter, equal to at least $9 ~\rm fm$, so that one can assume that collisions will dominantly involve valence nucleons.

\section{Results and discussions}

In order to distinguish  the  SM with inter-nucleon correlations from mean-filed model, it is better to avoid the impact of the core and hence to consider observables depending mainly on the peripheral nature of valence nucleons distributions.
$\pi$ mesons are used as probes to explore peripheral properties.
The inelastic processes, giving rise to $\pi$ production, play a significant role only at beam energies close to 800 MeV/nucleon\cite{BUUpi3}. 
Therefore we need a higher energy to produce more mesons in order to amplify this effect in HIC.
For this, we consider larger impact parameters (9 fm, 11 fm) associated to a high incident beam energy (1 GeV/u).

\begin{figure} [h]
\includegraphics[width=9 cm]{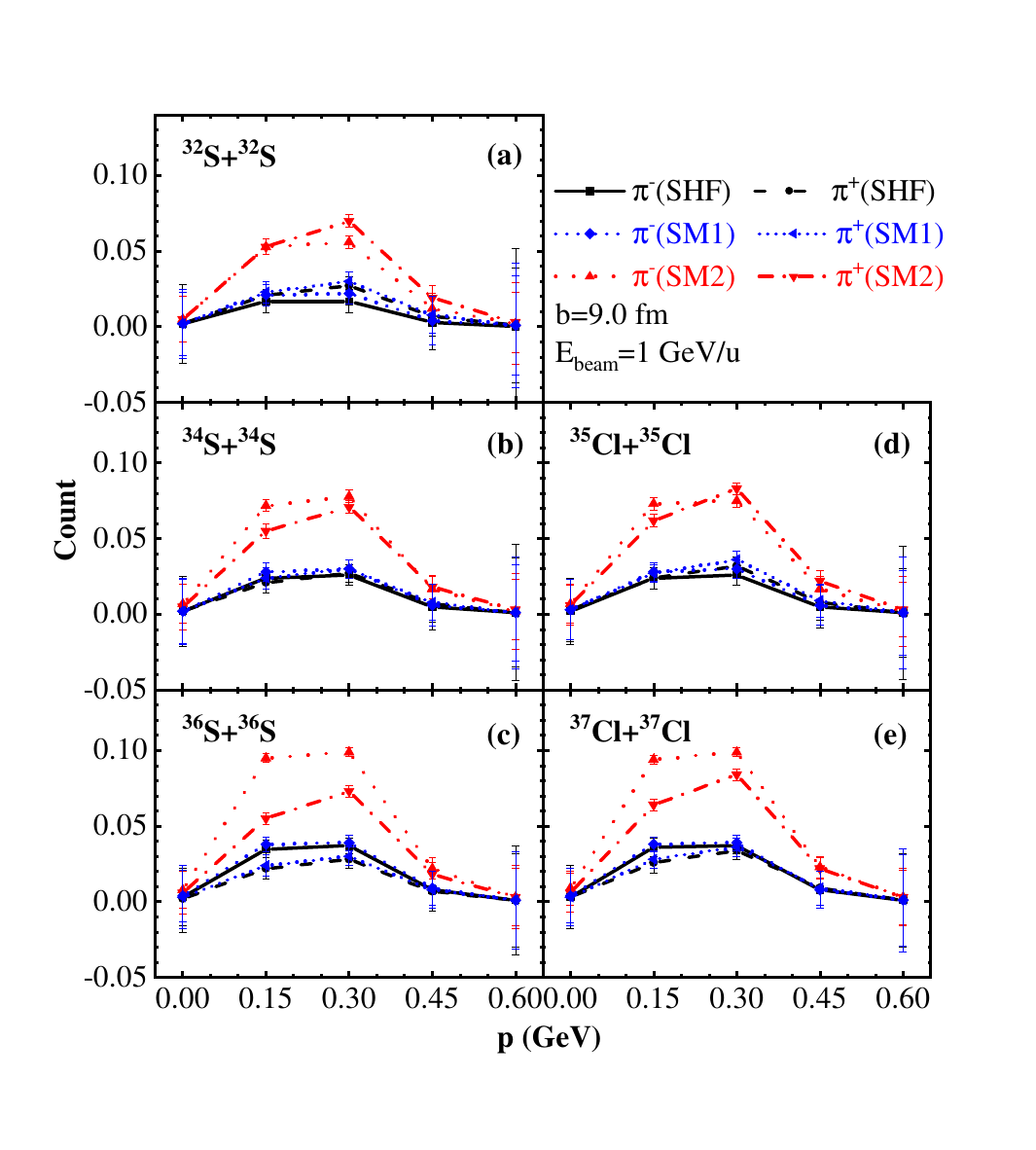}
\caption{ (Color online) Yields of $\pi^{-}$ and $\pi^+$ mesons as a function of momentum using different density distributions in SHF and USDB-SM in the $\rm^{A}S+\rm{^{A}S}$
    and in the $\rm ^{A}Cl+ \rm {^{A}Cl}$ reactions.
    One considers sulfur isotopes with $A=32$ (a), $A=34$ (b) and $A=36$ (c) nucleons in the $\rm^{A}S+\rm{^{A}S}$ reaction.
    For the $\rm ^{A}Cl+ \rm {^{A}Cl}$ reaction, one considers chlorine isotopes with $A=35$ (e) and $A=37$ (f).
    The used impact parameter and incident beam energy are 9 fm and 1 GeV/u, respectively.}\label{fig:count_p_SCL}
\end{figure}

The calculated yields of $\pi^{-}$ ,$\pi^+$ are depicted on Fig.\ref{fig:count_p_SCL} as a function of momentum with different initializations of valence nucleon profiles in the SHF, SM1 and SM2 for $\rm^{A}S+\rm{^{A}S}$ and  $\rm ^{A}Cl+ \rm {^{A}Cl}$ reactions.
Almost identical $\pi$ distributions are obtained with SHF and SM1, which is due to their similar nucleon density distributions when radii is around 4.5 fm.
Conversely, a large difference between SHF and SM2 in $\pi^{-}$ and $\pi^+$ yields emerges in the results, which is induced by the different nucleon density distributions arising from SM2 and SHF.
The produced $\pi$ meson yields are much larger with SM2 and are almost twice as large as those provided by SHF under the same entrance channel conditions.
The valence nucleon density of SM2 is also more important than that of SHF when radii are larger than 4.5 fm (see Fig.\ref{fig:init_S}).
Moreover, the yields of $\pi^{-}$, $\pi^+$ share a common peak at p=0.15 - 0.30 GeV, where the largest differences between SHF and SM2 results occur as well.
For reactions involving both sulphur and chlorine isotopes, the yields of $\pi^-$ mesons increase along with the number of neutrons of considered isotopes, whereas the yields of $\pi^+$ mesons are almost constant.
This arises because the $\pi^-$ mesons are mostly produced from neutron-neutron collisions, whereas the $\pi^+$ mesons are dominantly generated by proton-proton collisions.

\begin{figure}[h]
\includegraphics[width=9 cm]{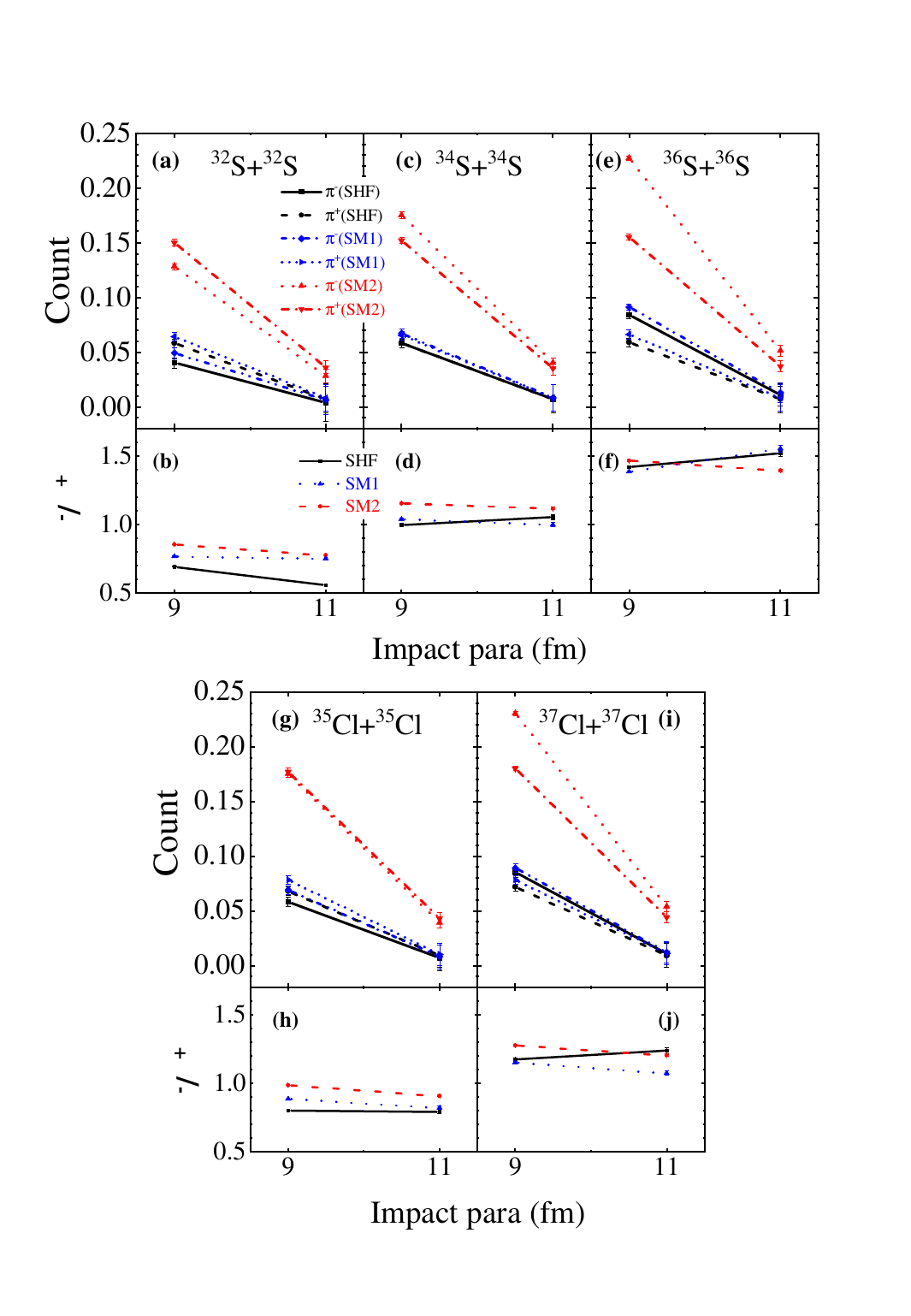}
\caption{\label{fig:count_A_all} (Color online) The yields of $\pi^{-}$ ,$\pi^+$ and the $\pi^{-}/\pi^+$ ratio as a function of impact parameter b with different nucleon density distribution in the SHF , SM1 and SM2 in the reactions of $\rm ^{A}S+ \rm {^{A}S}$ whose A=32 (a)-(b), A=34 (c)-(d),  A=36 (e)-(f)  and  $\rm ^{A}Cl+ \rm {^{A}Cl}$ whose A=35 (g)-(h), A=34 (i)-(j) at an incident beam energy of 1 GeV/nucleon.}
\end{figure}

Fig.\ref{fig:count_A_all}(a),(c),(e),(g),(i) show the yields of $\pi^{-}$ ,$\pi^+$  as a functions of impact parameter b with different nucleon density distribution in the SHF , SM1 and SM2 in different reactions.
One can see that the $\pi$ yield decreases sharply when the impact parameter is increased to 11 fm. 
The $\pi$ yields of all reactions are noticeably reduced by 25-30\% compared to those obtained at a impact parameter of 9 fm.
This arises because fewer nucleons are involved in the collision at larger impact parameters.

We now consider the $\pi^-/\pi^+$ ratio in the reactions considered above (see Fig.\ref{fig:count_A_all}(b),(d),(f),(h),(j)).
The $\pi^-/\pi^+$ ratio increases with the mass number of the colliding system in the considered reaction, because their number of protons remains the same, while their number of neutrons increases.
Consequently, more $\pi^-$ mesons are produced in a heavier system, so that the $\pi^-/\pi^+$ ratio augments.
As the impact parameters change, the $\pi^-/\pi^+$  ratio also changes.
This makes it possible to distinguish between the results arising from the SHF and SM1 calculations.
Indeed, The $\pi^-/\pi^+$ ratio increases(decreases) with the $\rm \rho_n/\rho_p$ increases(decreases)(see Fig.\ref{fig:npratio_r}).
When the impact parameter comes to 11 fm,  one should consider the $\rm \rho_n/\rho_p$ at around 5.5 fm.
At this point, comparing SHF and SM1, there is a significant difference in the  $\rm \rho_n/\rho_p$ ratios of $\rm ^{32}S$, $\rm ^{36}S$ and $\rm ^{37}Cl$.
Therefore, The $\pi^-/\pi^+$ ratios in the $\rm ^{32}S+ \rm {^{32}S}$, $\rm ^{36}S+ \rm {^{36}S}$ and $\rm ^{37}Cl+ \rm {^{37}Cl}$ reaction make a difference between SHF and SM1.
The double $\pi^-/\pi^+$ ratio approach \cite{BUUpi, DR, DR2, DR3, DR4, DR5} can also be used to differentiate between the SHF and SM1 models:
\begin{equation} \label{eq6}
Dr = \frac{(\pi^-/\pi^+)_{\rm ^{37}Cl+ \rm ^{37}Cl}}{(\pi^-/\pi^+)_{\rm ^{32}S+ \rm {^{32}S}}}.
\end{equation}
where $Dr$ is the value of the double $\pi^-/\pi^+$ ratio. The calculated values are shown in the Table.\ref{tabDr}.

\begin{table}[h]
\renewcommand{\arraystretch}{1.2}
        \caption{Double $\pi^-/\pi^+$ ratio.}
\begin{tabular}{@{}lllll@{}}
\toprule
~~~~~b   & SHF   & SM1   & SM2      \\
\toprule
~~~~9 fm ~~~~~~~  & 1.698~~~~~~~ & 1.501~~~~~~~ & 1.494~~~ \\
\hline
~~~11 fm  & 2.227 & 1.431 & 1.552  \\
\toprule

\end{tabular}
\label{tabDr}
\end{table}

One can see from the Table.\ref{tabDr}, for different impact parameters, the calculated double $\pi^-/\pi^+$ ratios of USDB-SM are quite similar, approximately equal to 1.5.
This due to a smoother change in the $\rm \rho_n/\rho_p$ ratio of the tail of density distribution, reflecting the inter-nucleon correlations effect.
However, $Dr(\rm SHF)$ is greater than this value, and which increases quickly with increasing impact parameter.
It is found that, with the impact parameter of 11 fm, the $Dr$ varies by a factor about 1.56 in the SHF and USDB-SM.
This large difference indicates that the $Dr$ is a proper observable to distinguish between USDB-SM and SHF. 
This is worth exploring from an experimental point of view.

\section{Summary}

$\pi$ productions in HIC depend on the nucleon density distribution of the colliding nuclei in heavy ion collisions.
Nucleon densities are typically generated in a mean-field framework. 
However, inter-nucleon correlations may play an important role in nucleon distribution in nuclei.
We have assessed the effect of valence nucleon densities of the initial colliding nuclei in SHF and USDB-SM approaches on $\pi$ yields in this work.

We considered colliding systems of sulfur and chlorine isotopes ($\rm ^{A}S+ \rm {^{A}S}$, $\rm ^{A}Cl+ \rm {^{A}Cl}$) in our applications pertaining to peripheral collisions.
We noticed that $\pi$ yields are twice as large when considering inter-nucleon correlations with the harmonic oscillator length $b_{HO}$ = 2.5 fm (SM2), because densities are enhanced in the surface region, and this augmentation acts similarly on $\pi^+$ and $\pi^-$ meson productions.
The $\pi^-/\pi^+$ ratio varies significantly with nucleon number. 
The changes in the $\pi^-/\pi^+$ ratio caused by the impact parameters are consistent with the variations in the density ratio of nuclei.
By investigating the two reactions $\rm ^{37}Cl+ \rm ^{37}Cl$ and $\rm ^{32}S+ \rm ^{32}S$, which differ most significantly at 11 fm,
we found that the double ratio of $\pi^-/\pi^+$ is a proper observable to distinguish between SM1 and SHF, for which $Dr(\rm{SHF})$/$Dr(\rm{SM1}) \simeq 1.56$ (see Eq.(\ref{eq6})). 
Moreover, the double $\pi^{-}/\pi^{+}$ ratios of SM change smoother than those of SHF. 

Experimental $\pi$ yields analysis would provide with an interesting perspective about the study of the effect of valence nucleon densities in HIC.
One could obtain information about the inter-nucleon correlations induced by the nuclear force, which is neglected when using Thomas-Fermi or SHF nucleon densities.
Large differences have been noticed to occur in $\pi$ yields when nucleon densities are taken from either SHF or USDB-SM calculations.
Thus, we suggest that further experimental and theoretical studies should be made to confirm or infirm the dependence of inter-nucleon correlations on the $\pi$ production process occurring in heavy ion collisions.

\section{Acknowledgements}

Prof.~Gao-Chan Yong is greatly thanked for his advice.
This work is supported in part by the National Natural Science
Foundation of China under Grant Nos. 11775275, 11435014.

\end{document}